\documentclass{amsart}

\usepackage{epsfig}
\usepackage{graphicx}
\usepackage{amsmath,amssymb,amsfonts,latexsym}
\usepackage{graphicx}

\usepackage{amssymb}
\usepackage{enumerate, xspace}
\usepackage{dsfont}
\usepackage{afterpage}
\usepackage{changebar}
\usepackage{amsthm}
\usepackage{rotating}

\numberwithin{equation}{section}

\begin{document}

\title[Wavefunction transformation]
{Wavefunction transformation due to changing group and phase velocities: an expos${\rm \acute{E}}$}
\author {Andrew Das Arulsamy}
\address{Condensed Matter Group, Division of Interdisciplinary Science, F-02-08 Ketumbar Hill, Jalan Ketumbar, 56100 Kuala-Lumpur, Malaysia}
\email{sadwerdna@gmail.com}

\keywords{Geometric and dynamic phases; Phase and group velocities; Wavefunction transformation}

\date{\today}

\begin{abstract}
If a time-dependent Hamiltonian is allowed to evolve adiabatically, and if it returns to its original form, then the ground state wavefunction must have picked up the dynamic or(and) the geometric phase factor(s) due to some interaction during the above evolution. Here, (i) we invoke the Pancharatnam's notion of phase retardation to show that the microscopic origin of Berry's phase is due to a change in the phase velocity (of a wavefunction), and (ii) the dynamic phase factor is shown to have its origin in the group velocity of a wavefunction. We also expose that the emitted photon's polarization during an electronic transition has its origin in the changes of the phase velocity of the electron wavefunction. Here, we prove the above statements, (i) and (ii), such that they are not due to some interpretational issues, hence, an expos${\rm \acute{e}}$. 
\\ \\
PACS: 03.65.Vf
\end{abstract}

\maketitle

\section{Introduction}

Geometric and dynamic phases, phase and group velocities are the primary variables that will come to play in the following sections. Our motivation here is to show microscopically (within the proper rules of quantum mechanics) why and how the phase and group velocities can be properly related to the geometric and dynamic phases, respectively, including their physical implications. Geometric phase has been ``anticipated'' much earlier (in 1938) by Rytov and Vladimirskii~\cite{tov,tov2}, which was first formally proven to exist by Pancharatnam using the Poincar${\rm \acute{e}}$ sphere~\cite{panch}. Pancharatnam proved that the geometric phase is responsible for the changes in the coherent light intensity such that the light intensity can be associated to a change in the phase polarization between two coherent light beams, which are in the same polarization states~\cite{panch,panch2}. We will dig deep into his work shortly. For now, the readers are referred to Refs.~\cite{ber,ber3,ber4,sesh,sesh2} for complete historical perspectives on the geometric phase. Subsequently, Berry~\cite{ber2} independently obtained the formula to calculate geometric phase for cyclic and adiabatic quantum mechanical processes that eventually explains the Aharonov-Bohm effect~\cite{aha}, which also, turns out to be very important in quantum physics because the changes in geometric phase are observable by means of polarized light intensity~\cite{panch} and applied magnetic field~\cite{aha}. 

Apart from Pancharatnam, Berry's phase has also been shown to exist for noncyclic quantum processes by Samuel and Bhandari~\cite{bhan}, while Wilczek and Zee have exposed that the said phase cannot be gauged away (gauge invariant) for an adiabatic process with degenerate energy levels~\cite{zee}. Furthermore, many studies were carried out to confirm the ubiquitous property of the Berry's phase, namely, (a) Aharonov and Anandan have generalized the said phase for nonadiabatic processes with cyclic eigenstate~\cite{ahan}, (b) Hannay have derived the classical representation of Berry's phase known as the Hannay angle~\cite{han}, and (c) Simon proved the mathematical correspondence between the geometric phase and topology using the notion of fibre bundle~\cite{simon}. Some of these recent investigations are well documented in Ref.~\cite{roh}, and also in a thesis (and references therein) by Durstberger~\cite{kath}. 

Given these background, our main contributions here are to show (i) the existence of time-dependent phase change in the geometric phase such that the phase or(and) group acceleration(s) is(are) responsible for the time-dependent phase change, which gives rise to nonzero Pancharatnam-Berry and dynamic phases, and (ii) the origin of photon polarization is shown to be due to the changes in the phase velocity involving electronic transitions between two orthonormalized energy levels. Points (i) and (ii) expose the physical origin for the sentence ``a wavefunction picks up or drops a phase factor'', which can be related to the so-called wavefunction transformation. The detailed motivations of our work can be understood from the following expositions.

\subsection{Berry's phase}  

The time($t$)-dependent Schr${\rm\ddot{o}}$dinger equation~\cite{david}, 
\begin {eqnarray}
i\hbar\frac{\partial}{\partial t}\Psi(t) = H(t)\Psi(t) = E(t)\Psi(t). \label{eq:1}
\end {eqnarray}  
Here $\Psi(t)$ and $E(t)$ are the $t$-dependent eigenfunction (wavefunction) and eigenvalue (energy), respectively. The wave functions are assumed to be orthonormalized ($\cdots, n, m, \cdots$) such that, $\langle\Psi_n(t)|\Psi_m(t)\rangle = \delta_{nm}$, where $\delta_{nm}$ is the usual Kronecker delta with $\{n,m\} \in \mathbb{N}^*$, $\mathbb{N}^*$ is the set of natural numbers excluding zero and~\cite{david} 
\begin {eqnarray}
\Psi_n(t) = \sum_nc_n(t)\psi_m(t)e^{i\theta_n(t)}, \label{eq:3} \\
\theta_n(t) = -\frac{1}{\hbar}\int_0^tE_n(t'){\rm d}t'. \label{eq:4}
\end {eqnarray}  
The complete orthogonal set can be defined as $\{\cdots, \Psi_n(t), \Psi_m(t), \cdots\}$, $e^{i\theta_n(t)}$ denotes the dynamic phase factor, and $c_n(t)$ is the $n^{\rm th}$ eigenstate ($|\Psi_n(t)\rangle$) coefficient where $|c_n(t)|^2 + |c_m(t)|^2 = 1$ for a two-level ($n$ and $m$) system. For nondegenerate energy levels, the exact coefficient~\cite{david}, 
\begin {eqnarray}
\dot{c}_m(t) = -c_m(t)\langle\psi_m(t)|\dot{\psi}_m(t)\rangle - \sum_{n \neq m}c_n(t)\frac{\langle\psi_m(t)|\dot{H}(t)|\psi_m(t)\rangle}{E_n(t) - E_m(t)}e^{i(\theta_n(t) - \theta_m(t))}, \nonumber \\ \label{eq:5}
\end {eqnarray}  
where a single dot (say, $\dot{x}$) carried by a $t$-dependent variable means appropriate differentiation has been carried out with respect to time. If the Hamiltonian evolves adiabatically, Eq.~(\ref{eq:5}) reduces to $\dot{c}_m(t) = -c_m(t)\langle\psi_m(t)|\dot{\psi}_m(t)\rangle$ and therefore~\cite{david}, 
\begin {eqnarray}
c_m(t) = c_m(0)e^{i\gamma_m(t)}, \label{eq:6} \\
\gamma_m(t) = i\int_0^t\bigg\langle\psi_m(t')\bigg|\frac{\partial}{\partial t'}\psi(t')\bigg\rangle {\rm d}t', \label{eq:7}
\end {eqnarray}  
that gives (from Eq.~(\ref{eq:3}))
\begin {eqnarray}
\Psi'_n(t) = c_n(0)e^{i\theta_n(t)}e^{i\gamma_n(t)}\psi_n(t). \label{eq:8}
\end {eqnarray}  
Here, $e^{i\gamma_n(t)}$ is the geometric phase factor, $c_n(0) = 1$, and note (from Eqs.~(\ref{eq:3}) and~(\ref{eq:8})) $\langle\Psi_n(t)|\Psi_n(t)\rangle = \langle\Psi'_n(t)|\Psi'_n(t)\rangle = \langle\psi_n(t)|\psi_n(t)\rangle$ due to adiabatic evolution. According to Berry~\cite{ber2}, if the $t$-dependence in $\psi(t)$ is due to some other $t$-dependent parameter, say $X(t)$ such that $\psi(t) \rightarrow \psi[X(t)]$, then~\cite{ber2}  
\begin {eqnarray}
\gamma_n(\textbf{X}(t)) = i\int_{\textbf{X}(t_0)}^{\textbf{X}(t_1)}\langle\psi_n(\textbf{X}(t))|\nabla\psi_n(\textbf{X}(t))\rangle {\rm d}\textbf{X}(t), \label{eq:9} \\ \gamma_n(T) = i\oint\langle\psi_n(\textbf{X})|\nabla\psi_n(\textbf{X})\rangle {\rm d}\textbf{X}, \label{eq:10}
\end {eqnarray}  
where the time notation $t_0 < t_1$, while $T$ is the time taken for the Hamiltonian to return to its original form. Apparently, Eqs.~(\ref{eq:9}) and~(\ref{eq:10}) are nonintegrable and these integrals cannot be zero if $\psi(t) \rightarrow \psi[X_1(t),X_2(t),\cdots,X_N(t)] = \psi(\textbf{X})$, $N > 1$ and at least, $X_1(t) \neq X_2(t)$. Here, $\gamma(T)$ is known as the Berry's phase. 

\subsection{Pancharatnam's phase}

Pancharatnam's phase~\cite{panch} is associated to the phase advance of one polarized beam compared to another beam such that the intensity of these two combined beams are maximum only if the polarized beam ($I_1$) with a phase advance ($\delta$) has its phase retarded by the same amount, compared to the second beam ($I_2$). These two beams are not necessarily in the same polarization state. The resultant intensity~\cite{panch},
\begin {eqnarray}
I = I_1 + I_2 + 2\sqrt{I_1I_2\cos^2{[(1/2)c]}}\cos{\delta}, \label{eq:11}
\end {eqnarray}  
where $\cos^2{[(1/2)c]}$ is known as the Pancharatnam's similarity factor, and $c$ denotes the angular separation between the two beams due to different polarization states. Maximum $I$ is obtained if the two beams are identically polarized ($c = 0$), and there is no phase difference between the beams ($\delta = 0$). For example, maximum intensity, $I = 4I_1$ can be obtained if $I_1 = I_2$, however, any change to the intensity, independent of the phase retardation, is independent of $\delta$. In other words, any change in the resultant intensity, cannot be attributed solely to the change in the phase difference. Here, the phase retardation notion introduced by Pancharatnam refers to the mechanism of altering the phase velocity ($v_p(t)$), which will be used to show that Eqs.~(\ref{eq:9}) and~(\ref{eq:10}) are nothing but equations that record the changes in the phase and group momenta (after multiplying with $\hbar/i$) due to the changes in the phase and group velocities of a wavefunction. On the other hand, Eq.~(\ref{eq:4}) refers to a sole change in the group momentum due to a change in the group velocity ($v_g(t)$) of a wavefunction. The $t$-dependences for both the phase and group velocities require a form of wavefunction transformation that reads, 
\begin {eqnarray}
\psi(t) \rightarrow \psi[v_p(t),v_g(t)] = \psi(\textbf{X}), \label{eq:12}
\end {eqnarray}  
where $N = 2$, and $N$ is the number of $t$-dependent parameters ($v_p(t)$ and $v_g(t)$) responsible for the $t$-dependence of a wavefunction. The irony here is that, contrary to the claims made in Refs.~\cite{ber,ber3,ber4}, we actually need to invoke the Pancharatnam's notion of phase retardation (changing phase velocity due to an interaction between a polarizer and a pencil beam) to expose the physical origin of Berry's phase. 

However, Rytov~\cite{tov} and Vladimirskii~\cite{tov2} did come close to the Pancharatnam's notion with the remark---different phase velocities are responsible for the polarization rotation of an outgoing beam, relative to an incident beam. In contrast, Pancharatnam proved his notion of phase velocity retardation by associating the resultant intensity ($I$) to $\delta$ (see Eq.~(\ref{eq:11})). He further proved that a beam of polarization $C$ can be decomposed into two beams of polarizations $A$ and $B$ such that $C$ and $A$, and $C$ and $B$ are in phase, but $A$ and $B$ may not be in phase if $\delta_{A} \neq \delta_{B}$ where the Pancharatnam geometric phase factor is given by $ e^{i\delta} = e^{i[\pi - (1/2)\Omega_{ACB}]}$ and $\Omega_{ACB}$ is the angle of a geodesic triangle on a poincar${\rm \acute{e}}$ sphere~\cite{panch}. You can also refer to Fig.~13.09 in Ref.~\cite{smy} to observe the elliptic polarization of light ($\delta \neq 0$) in Cartesian coordinates (for a simpler visualization). 

\section{Wavefunction transformation due to changing phase and group momenta}

The wavefunction transformation, $\psi(t) \rightarrow \psi[X_1(t),X_2(t),\cdots,X_N(t)] = \psi(\textbf{X}(t))$ strictly means that $\gamma(t) \rightarrow \gamma(\textbf{X}(t))$ such that $\gamma(\textbf{X}(t)) = \gamma[v_p(t),v_g(t)]$, while the dynamic phase factor, $\theta(t) = \theta[v_p(t)]$ remains the same where $\gamma[v_p(t),v_g(t)] \neq \theta[v_g(t)]$. To prove this, we will first need to recall Eq.~({\ref{eq:7}}), which can be written as (using Eq.~(\ref{eq:1}))
\begin {eqnarray}
\gamma(t) &=& i\int_0^t\bigg\langle\psi(t')\bigg|\frac{\partial}{\partial t'}\psi(t')\bigg\rangle {\rm d}t' \label{eq:13} \\&=& \frac{1}{\hbar}\int_0^t\langle\psi(t')|E(t')|\psi(t')\rangle {\rm d}t'. \label{eq:14}
\end {eqnarray}  
Equation~(\ref{eq:14}) implies $\gamma(t)$ can be made to approach $\theta(t)$ such that $\gamma(t) \rightarrow \theta(t)$, which will be exposed shortly. Anyway, if we now transform the wavefunction, $\psi(t) \rightarrow \psi_{\rm new}(t) = e^{ig(t)}\psi(t)$, then Eq.~({\ref{eq:13}}) reads
\begin {eqnarray}
\gamma_{\rm new}(t) &=& -\int_0^t\bigg\langle\psi_{\rm new}(t')\bigg|\frac{\partial g(t')}{\partial t'}\bigg|\psi_{\rm new}(t')\bigg\rangle {\rm d}t' + i\int_0^t\bigg\langle\psi_{\rm new}(t')\bigg|\frac{\partial}{\partial t'}\psi_{\rm new}(t')\bigg\rangle {\rm d}t' \nonumber \\&=& \gamma[v_p(t)] + \gamma[v_g(t)], \label{eq:15}
\end {eqnarray}  
therefore, $\gamma(t) \rightarrow \gamma[v_p(t)] + \gamma[v_g(t)] = \gamma_{\rm new}[v_p(t),v_g(t)]$ where the origin for the change in the phase velocity is due to the first integral in Eq.~({\ref{eq:15}}), while the second integral in Eq.~({\ref{eq:15}}) determines the changes in the group velocity. It should be obvious here that the above velocity changes do not necessarily mean $v_p(0) \neq v_p(t_1)$ and/or $v_g(0) \neq v_g(t_1)$ for both $t - 0 = T$ and $t - 0 \neq T$ (recall that $T$ is the time taken for the Hamiltonian to return to its original form). You should also be aware here that the changes in the group and phase velocities exist due to some interaction between time, $0$ and $t$, and between an electron and an external ``force''. In fact, this interaction gives rise to the wavefunction transformation such that the original wavefunction picks up (or drops) the relevant phase factors accordingly. The above interaction is the root cause for this transformation or responsible for the wavefunction to pick up (or drop) a phase factor. On the other hand, for nonadiabatic processes, one can rewrite Eq.~({\ref{eq:5}})
\begin {eqnarray}
\dot{c}_m(t) = -c_m(t)\langle\psi_m(t)|\dot{\psi}_m(t)\rangle - \frac{\partial \Gamma(t)}{\partial t}, \label{eq:5a}
\end {eqnarray}  
and its solution, 
\begin {eqnarray}
c_m(t) = c_m(0)e^{i\gamma_m(t)} - \Gamma(t), \label{eq:6a}
\end {eqnarray}  
where $\partial\Gamma(t)/\partial t$ symbolically denotes the second term on the right-hand side of Eq.~({\ref{eq:5}}). Subsequently, using Eq.~({\ref{eq:3}}), one obtains
\begin {eqnarray}
\Psi''_m(t) = \big[e^{i\gamma_m(t)} - \Gamma(t)\big]\psi_m(t)e^{i\theta_m(t)}, \label{eq:3a}
\end {eqnarray}  
in which, the geometric phase factor remains intact (see Eq.~({\ref{eq:3a}})) while the original wavefunction has picked up another complicated function ($\Gamma(t)$) as it should be for nonadiabatic processes.   

The above proof shows why $\gamma_{\rm new}(t)$ is responsible for both $v_p(t)$ and $v_g(t)$, but to understand why $\gamma_{\rm new}(t)$ is also responsible for $v_g(t)$, we do the following---if we now write $\psi(t) \rightarrow \psi_{\rm new}'(0) = e^{-iEt/\hbar}\psi$ where $g(t) = -Et/\hbar$ and $\psi$ is $t$-independent, and if we now switch-on whatever $t$-dependent interaction (between $0$ and $t$) required to allow $\psi_{\rm new}'(0)$ to evolve, then (using Eq.~({\ref{eq:13}})),
\begin {eqnarray}
\gamma'(t) \rightarrow \theta'(t) = \frac{E}{\hbar}\int_0^t{\rm d}t' = \frac{Et}{\hbar}. \label{eq:16}
\end {eqnarray}  
After substituting $E(t) = E$ ($t$-independent energy) in Eq.~(\ref{eq:4}), one obtains $\theta(t) = -Et/\hbar$, which is nothing but the dynamic phase factor introduced by-hand in $\psi_{\rm new}'(0) = e^{-iEt/\hbar}\psi$. Hence, we now know why $\gamma'(t) \rightarrow \theta'(t)$---the introduced phase factor is a dynamic phase factor. After the interaction, $\psi_{\rm new}'(t) = e^{-iEt/\hbar}\psi e^{iEt/\hbar} = \psi$, which signifies a group acceleration or deceleration has taken place such that $v_g(0) \rightarrow v_g(t)$ where both $v_g(0)$ and $v_g(t)$ cannot be zero, and these velocities do satisfy $v_g(0) \neq v_g(t)$ even though $E$ is $t$-independent because $E$ is determined by $\psi$ alone, and not by some phase factors (dynamic or geometric). Therefore, indeed the phase factor, $e^{-iEt/\hbar}$ gives rise to the dynamic phase factor, as well as to the changes in the group velocity (see Eqs.~(\ref{eq:16}) and~(\ref{eq:4})) during the interaction time (such that $v_g(0) \neq v_g(t)$). This means that, we need a proper $t$-dependent wavefunction (recall $\psi_{\rm new}(t) = e^{ig(t)}\psi(t)$) to allow variations in both the phase and group velocities (see Eq.~(\ref{eq:15})). If we use an improper wavefunction, namely, $\psi^{\rm improper}_{\rm new}(t) = \psi_{\rm a}(t)\psi_{\rm b}(t)$, then (using Eq.~(\ref{eq:13})) 
\begin {eqnarray}
\gamma^{\rm improper}_{\rm new}(t) &=& i\int_0^t\bigg\langle\psi^{\rm improper}_{\rm new}(t')\bigg|\frac{\partial \psi_{\rm a}(t')}{\partial t'}\psi_{\rm b}(t')\bigg\rangle {\rm d}t' \nonumber \\&& + ~i\int_0^t\bigg\langle\psi^{\rm improper}_{\rm new}(t')\bigg|\psi_{\rm a}(t')\frac{\partial \psi_{\rm b}(t')}{\partial t'}\bigg\rangle {\rm d}t' \label{eq:15a} \\&=&
\gamma[v_g(t)]. \label{eq:15aa}
\end {eqnarray}  
As expected, wavefunctions with $t$-independent phase factors do not give rise to changing phase velocity, and therefore, $v_p(t)$ is always a constant for such cases (physical systems with improper wavefunctions). In addition, note that if we use $\psi^{\rm improper}_{\rm new}(t) = \psi_{\rm a}\psi_{\rm b}(t)$ ($\psi_{\rm a}$ is $t$-independent), then the first term on the right-hand side of Eq.~(\ref{eq:15a}) is zero.   

The following example may explain the proper wavefunction transformation needed to capture the phase velocity. For example, $\psi(t) \rightarrow \psi_{\rm new}(t) = e^{ig(t)}\psi(t)$ and/or $\psi'_n(\textbf{r} - \textbf{R}) \rightarrow \psi_n(\textbf{r},\textbf{r} - \textbf{R})$ where the former wavefunction ($\psi(t)$) picks up a phase factor, while the second wavefunction ($\psi'_n(\textbf{r} - \textbf{R})$) picks up a new variable, $\textbf{r}$ (see the example below). Example---consider the Schr${\rm \ddot{o}}$dinger equation that captures the motion of an electron confined by a potential, $V(\textbf{r}-\textbf{R})$ through a region where $\nabla \times \textbf{A} = \textbf{B} = 0$ such that the vector potential, $\textbf{A} \neq 0$, is given by~\cite{ber2,david}
\begin {eqnarray}
&&\bigg[\frac{1}{2m_e}\bigg(\frac{\hbar}{i}\nabla - e\textbf{A}(\textbf{r})\bigg)^2 + V(\textbf{r}-\textbf{R})\bigg]\psi_n(\textbf{r},\textbf{r} - \textbf{R}) = E_n\psi_n(\textbf{r},\textbf{r} - \textbf{R}). \label{eq:17}
\end {eqnarray}  
Here $m_e$ is the electron mass, $e$ is the electron charge, $V(\textbf{r}-\textbf{R})$ is the potential experience by the electron where $|\textbf{r}-\textbf{R}|$ denotes the displacement of the confined electron, while $\textbf{R}$ and $\textbf{r}$ denote the distances between a confined electron and the magnetic field, \textbf{B}, before ($\textbf{R}(0)$) and after ($\textbf{r}(t)$) the electron-displacement, respectively. Considering a proper wavefunction~\cite{david}, $\psi(\textbf{r},\textbf{r} - \textbf{R}) = e^{ig(\textbf{r})}\psi'(\textbf{r} - \textbf{R})$, we can rewrite Eq.~(\ref{eq:17}),
\begin {eqnarray}
\bigg[-\frac{\hbar^2}{2m_e}\nabla^2 + V(\textbf{r}-\textbf{R})\bigg]\psi'_n(\textbf{r} - \textbf{R}) = E_n\psi'_n(\textbf{r} - \textbf{R}), \label{eq:18}
\end {eqnarray}  
where
\begin {eqnarray}
g(\textbf{r}) &=& \frac{e}{\hbar}\int_{\textbf{R}(0)}^{\textbf{r}(t)}\textbf{A}(\textbf{r}'(t)){\rm d}\textbf{r}'(t), \label{eq:19} 
\end {eqnarray}
such that Eq.~(\ref{eq:18}) implies $\psi' = \psi'(\textbf{r}-\textbf{R})$ because $\psi'$ is independent of any effect from $\textbf{B}$, whereas $\psi = \psi(\textbf{r},\textbf{r}-\textbf{R})$ is due to some nonzero vector potential, $\textbf{A}(\textbf{r})$. Now, to find the phase and group momenta, we multiply the integrand in Eq.~(\ref{eq:10}) by $\hbar/i$ to obtain
\begin {eqnarray}
\frac{\hbar}{i}\langle\psi_n(\textbf{X})|\nabla_{\textbf{X}}\psi_n(\textbf{X})\rangle &=& \frac{\hbar}{i}\langle\psi_n(\textbf{r}, \textbf{r} - \textbf{R})|\nabla_{\textbf{r}}\psi_n(\textbf{r}, \textbf{r} - \textbf{R})\rangle \nonumber \\&=& -e\textbf{A}(\textbf{R}) - i\hbar\int[\psi'_n(\textbf{r}-\textbf{R})]^*\nabla_{\textbf{r}}[\psi'_n(\textbf{r}-\textbf{R})]{\rm d}^3\textbf{r}. \label{eq:20}
\end {eqnarray}  
where $\gamma'(t) \rightarrow \gamma_{\rm new}[v_p(t),v_g(t)]$ (from Eqs.~(\ref{eq:20}) and~(\ref{eq:15})) implies a wavefunction transformation that reads, $\psi(\textbf{r} - \textbf{R}) \rightarrow e^{ig(\textbf{r})}\psi'(\textbf{r} - \textbf{R}) = \psi[v_p(t),v_g(t)]$. In other words, Eq.~(\ref{eq:20}) is nothing but what has been proven earlier (see Eq.~(\ref{eq:15})), which also implies that the phase momentum ($e\textbf{A}(\textbf{R})$) is observable, while the group momentum is not, simultaneously that is (see the second term ($i\hbar \times$integral) in Eq~(\ref{eq:20})). Now this is cute, if you confine the electron inside a box, as Berry did~\cite{ber2}, and transport the box adiabatically, then $V(\textbf{r} - \textbf{R})$ is the potential that confines an electron in a box ($V(\textbf{r} - \textbf{R}) \rightarrow V(x,y,z)$), which in turn means $\psi'_n(\textbf{r} - \textbf{R}) \rightarrow \psi'_n(x,y,z)$ is a stationary state, and therefore, the integral is zero (the change in group momentum is zero). Experimentally, there is no such thing as, you confine an electron in a box, and transport the box adiabatically. Therefore, the microscopic origin of the observed Aharonov-Bohm effect~\cite{aha} is due to zero group momentum change of the electronic wavefunction (see Eq.~(\ref{eq:15})). Indeed, the nonzero phase momentum change gives rise to a geometric phase factor (known as the Aharonov-Bohm effect), which has been observed via the measurement of phase difference between the incoming and outgoing beams, after some interaction in a region where $\textbf{B} = 0$ but $\textbf{A} \neq 0$ and these beams never come in boxes~\cite{cham} meaning, $V(\textbf{r} - \textbf{R}) \neq V(x,y,z)$, as correctly formulated and observed by Pancharatnam with his pencil beams~\cite{panch}.

We now justify why Eq.~(\ref{eq:15}) gives rise to the photon polarization. The second irony here is that, contrary to the claim made in Refs.~\cite{ber3,ber4}---Berry's phase is quantum mechanical, while Pancharatnam's phase is for classical light beams---is technically and physically false. We expose here that the polarization state of light cannot be detached from the phase velocity of an electron (the so-called quantum particle) because the phase polarization of light (photons) is microscopically originated from the phase velocity change of an electron during some electronic transition. The photon gains, whatever the electron loses during an emission (stimulated or spontaneous). A common example in this respect is the existence of selection rules (with respect to the principal ($n$), magnetic ($m'$) and azimuthal ($l$) quantum numbers)---the electronic transition between two energy levels that correspond to two orthonormalized wave functions, $\psi_{n_1l_1m'}$ and $\psi_{n_2l_2m'}$ ($\langle\psi_{n_1l_1m'}|\psi_{n_2l_2m'}\rangle = 0$, $\Delta m' = 0, l_2 - l_1 = \pm1, n_2 - n_1 = 1$) determine the energy and the spin of an emitted photon~\cite{david}. Here, Eqs.~(\ref{eq:20}) and~(\ref{eq:15}) also imply that any change in the phase velocity during the above electronic transition, logically means the emitted photon gains some phase polarization. This effect can be observed experimentally by measuring the intensity of emitted photons (as originally proven by Pancharatnam~\cite{panch}), which exposes that the phase velocity ``lost'' by an electron during an electronic transition determines the polarization state of an emitted photon.                    

\section{Conclusions}

Our primary new result here is that we have proven that the phase and group velocities can be physically and mathematically associated to the geometric and dynamic phases acquired by a physical system through its wavefunction. In our proof, we also have shown mathematically that both the phase and group velocities of a wavefunction determine the geometric phase, while the group velocity of a wavefunction solely responsible for the dynamic phase. Along the way, we have noted that the origin of Berry's phase is due to the notion of phase retardation of coherent beams introduced by Pancharatnam, which is also applicable for quantum particles, namely electrons. For example, the origin of an emitted photon's polarization during an electronic transition between two orthonormalized energy levels depends on the changes in the phase velocity of that electron. Furthermore, using the notion of Pancharatnam's phase retardation, we have proven that the formula derived by Berry is nothing but the accumulation of both the phase and group momenta of a proper wavefunction. This means that the Pancharatnam's phase retardation is well defined and unambiguous for both light and quantum particles. Finally, we have defined the notion of wavefunction transformation as any changes to a wavefunction such that it picks up (or drops) a phase factor, or any variable completely due to some interaction.     
  
\section*{Acknowledgments}

This work was supported by Sebastiammal Innasimuthu, Arulsamy Innasimuthu, Amelia Das Anthony, Malcolm Anandraj and Kingston Kisshenraj. I am grateful to Alexander Jeffrey Hinde (The University of Sydney) for his unconditional help in providing some of the references.

\end{document}